\documentclass{iopart}

\usepackage{latexsym}
\usepackage{graphicx}
\usepackage{color}

\newcommand{\bra}[1]{\langle #1 |}
\newcommand{\ket}[1]{| #1 \rangle}
\newcommand{\braket}[2]{\langle #1 | #2 \rangle}
\newcommand{\mics}{\: \mu s}
\begin{document}

\title{On role of the atom-cavity detuning in bimodal cavity experiments}

\author{D Gon\c{t}a$^1$ and S Fritzsche$^{1,2}$}

\address{$^1$ Max-Planck-Institut f\"{u}r Kernphysik, Postfach 103980, \\
              \hspace*{0.28cm}D--69029 Heidelberg, Germany}

\address{$^2$ Physikalisches Institut der Universit\"{a}t Heidelberg,
              Philosophenweg 12,\\
          \hspace*{0.28cm}D-69120 Heidelberg, Germany}

\ead{gonta@physi.uni-heidelberg.de}
\ead{s.fritzsche@gsi.de}

\begin{abstract}
The coherent evolution of the atom-cavity state in bimodal (cavity)
experiments has been analyzed for a realistic time-dependence in
detuning the atomic transition frequency. Apart from a `smooth
switch' of the atomic resonance from one to the second mode of a
bimodal cavity, we considered also an additional (effective)
interaction between the field modes of the cavity, known as
`communication channel'. Comparison of our model computations has
been made especially with the measurements by Rauschenbeutel \etal
[2001 {\it Phys. Rev.} A {\bf 64} 050301] who demonstrated for the
first time the entanglement of the field modes in a bimodal cavity.
It is shown that the agreement between the (theoretically) predicted
and experimental phase shifts can be improved by allowing a
`communication' between the two field modes during a short but
finite switch of the atomic transition frequency from one mode to
the other. We therefore suggest that the details of the atom-cavity
detuning should be taken into account for the future interpretation
of bimodal cavity experiments.
\end{abstract}


\maketitle

\section{Introduction}

During the last decades, entanglement has been recognized as a key
feature of quantum mechanics that describes not only the correlation
in composite quantum systems but is useful also for applications.
Since the famous Bohr-Einstein debate and the seminal work of Einstein,
Podolsky and Rosen \cite{pr47} in 1935, indeed, a large number of
entanglement studies helped improve our present understanding of nonlocal
and nonclassical phenomena as they occur in the microscopic world. In
quantum engineering and quantum information theory \cite{nc},
moreover, entanglement has been found crucial for implementing new
(quantum) information protocols, such as super-dense coding
\cite{bew}, quantum cryptography \cite{eke}, or even simple quantum
algorithms \cite{gro}. However, despite of all progress in the design
and description of entangled quantum system, their manipulation and
controlled interaction with the environment still remains a great challenge
for experiment owing to the fragile nature of most quantum states. Among
various other implementations, an excellent control over the light fields
and atoms has been achieved especially by using neutral atoms that are
coupled to a high-finesse optical cavity \cite{haroche, apl90}.

In a recent experiment by Rauschenbeutel and coworkers \cite{pra64},
for example, the two modes of a superconducting (bimodal) cavity
were prepared in a maximally entangled state by using circular
Rydberg atoms. In this experiment, an entanglement of the two field
modes was achieved, by properly adjusting the detuning of the atomic
(transition) frequency of the Rydberg atom while it passes through
the bimodal cavity. The entangled state produced in the cavity is
probed later by a second atom that, after being detected, reveals
the coherent evolution of the superposition of the cavity states.
The coherence of the (two-qubit) cavity state was demonstrated by
varying the delay time after which the probe atom interacts with the
cavity. Different delay times give then rise to a oscillation in the
final-state probability for the second atom to be found in either
the \textit{ground} or \textit{excited} state of the two-level
configuration, as supposed for the Rydberg atom. For this
final-state probability, an analytical expression was derived by
assuming an \textit{idealized} time evolution of the quantum state
of both, the photon field and the atoms, throughout the periods of
atom-cavity interactions. Despite the high quality of the cavities
today, however, such an \textit{ideal} time evolution neglects a
number of relevant effects, including the relaxation of the cavity
field, the influence of external and internal stray fields, or
imperfections due to the cavity mirrors and cavity geometry. In
particular the proper matching of the atomic transition frequency to
(the frequency of) one or the other cavity mode (the so-called
atom-cavity detuning) plays an essential role on the superposition
of the cavity states, because the atom has to mediate the
interactions between the (cavity) modes while passing through the
cavity.

The experiments by Rauschenbeutel \etal \cite{pra64} nicely
demonstrated how the field modes of a bimodal cavity become
entangled or disentangled in a well controlled way by manipulating
the (de-) tuning of the atom-cavity interaction. In contrast to the
single-mode cavities, the use of bimodal cavities has been found an
important step towards the manipulation of complex quantum states
and for performing various fundamental tests in quantum theory
\cite{haroche1, pra70, pra76, pla339, pra68, jmo51, jmo52, pra66,
pra67}. For these bimodal cavities, its important to know how
(sensitive) the produced cavity state depends on the details and the
particular shape of the detuning process. Since the two nearly
degenerate frequencies of the cavity modes are fixed by the geometry
of the cavity, a detuning of the atom (transition) frequency with
regard to one of the field modes automatically affects also the
detuning with regard to the other mode. In our discussions below, we
shall often refer to this detuning of the atomic transition
frequency (of the Rydberg atoms passing through the cavity) with
regard to the field modes briefly as the \textit{atom-cavity
detuning}.

In the present work, we examine how the coherent evolution of the
cavity state of bimodal cavities, i.e.\ the superposition of the two
cavity modes, is affected by a realistic time-dependence of the
atom-cavity detuning and how the atoms, when passing through the
cavity, may interact with both cavity modes simultaneously. To
exhibit these effects, we make use of two models for the atom-cavity
detuning in which the (idealized) \textit{step-wise} detuning of the
atomic transition frequency from one to the other cavity mode is
replaced by (i) a smooth detuning that happens in a short but finite
(`switching') time interval and (ii) a simultaneous interaction of
the atom with both cavity modes leading to a wave mixing in the
cavity. We shall refer to these models as the (separate)
\textit{single-mode interaction} and \textit{communication channel}
model. In the latter model, the wave-mixing effects mainly arise due
to imperfections of the cavity mirrors that causes the cavity modes
to interact (communicate) with each other, cf.\ Section~2. To follow
the time evolution of the atom-cavity interaction, we combine
analytical solutions for the Jaynes-Cummings Hamiltonian with
numerical simulations, if the atom-cavity interaction is not
resonant. For both models, the single-mode interaction and
communication channel, the final-state probability for the `probe'
atom to be found in either the ground or excited state is then
compared with the experiments in \cite{pra64}. Using the cavity
parameters from these experiments, we show that more realistic
assumptions about the shape (and model) of the atom-cavity
interaction improves the agreement between the theoretical
predictions and experiment. A detailed account of the atom-cavity
detuning will therefore play an crucial role also for manipulating
and analyzing (more) complex cavity experiments in the future.

In Section~2 we start from the Jaynes-Cummings Hamiltonian to
describe the interaction of a two-level atom with the two field
modes of a bimodal cavity. For this, we first (re-) derive the
expression for the final-state probability in Subsection 2.1 that
the probe atom is detected in the excited state by using the matrix
formalism and by assuming an idealized \textit{step-wise} change in
the atom-cavity interaction. This derivation sets the framework for
introducing our more realistic models concerning the shape and
explicit form of the atom-cavity interaction in sections~2.2
(single-mode interaction model) and 2.3 (communication-channel). In
the latter model, we modify the Jaynes-Cummings Hamiltonian as to
allow the atom to interact with both cavity modes simultaneously. In
Section~3 then, the predictions from these models for the
final-state probability are compared with the data of \cite{pra64}.
In particular, here we display and discuss how the `switching time'
affects the outcome of the experiments. Finally, a few conclusions
are given in Section~4.

\section{Entanglement of two field modes using a bimodal cavity}

The use of the \textit{resonant} atom-cavity interaction regime is
the simplest way to generate entangled states between atoms and/or
the field modes of a cavity. For a sufficiently high enough quality
factor of the cavity mirrors, namely, this regime implies a `strong'
atom-field coupling for which the dissipation of field energy in
course of the atom-cavity evolution becomes negligible. Such a small
dissipation of the photon field plays a crucial role for
engineering of coherent states in the framework of cavity QED. Apart
from the quality of the mirrors, the correct matching of the atomic
transition frequency to the frequency of cavity field modes (the
so-called \textit{detuning}) is also an important ingredient in
order to achieve the resonant atom-cavity interaction regime.

In the following, let us first recall the basic components and
notions of the cavity QED experiments by Haroche and coworkers
\cite{haroche}. In these experiments, circular Rydberg atoms in
quantum states with principal quantum numbers 50 and 51 are treated
as `two-level' systems, being in the ground state $\ket{g}$ or the
exited state $\ket{e}$, respectively. Using the Stark shift
technique, the frequency of atomic $e \leftrightarrow g$ transition
can be tuned in a well controlled way to the nearly degenerate
frequencies of the two cavity field modes. If, for the moment, we
consider only a single-mode cavity, then the evolution of the
atom-cavity state is described (for both, a resonant and
non-resonant interaction of the atom with the cavity) by the
Jaynes-Cummings Hamiltonian \cite{jc}
\begin{eqnarray}\label{ham}
H & = & \hbar \omega_{0} S_{z} + \hbar \frac{\Omega}{2}
        \left( S_{+} a_1 + a_1^{+} S_{-} \right)
      + \hbar \omega_1 \left(a_1^{+} a_1 + \frac{1}{2} \right) \, ,
\end{eqnarray}
where $\omega_{0}$ is the atomic $e \leftrightarrow g$ transition
frequency, $\omega_1$ the frequency of the cavity field, and
$\Omega$ the atom-field coupling frequency. In the Hamiltonian
(\ref{ham}), moreover, $a_1$  and $a_1^+$ denote the annihilation
and creation operators for a photon in the cavity, acting upon the
Fock states $\ket{n}$, while $S_{-}$ and $S_{+}$ are the spin
lowering and raising operators that act upon the atomic states
$\ket{e}$ and $\ket{g}$, which are the `eigenstates' of the spin
operator $S_z$ with eigenvalues $+1/2$ and $-1/2$, respectively.
Furthermore, if there is not more than \textit{one} photon in the
cavity, the overall atom-field state for a resonant atom-cavity
interaction, i.e.\ for a zero detuning ($0 = \omega_0 - \omega_1$),
evolves according to \cite{puri}
\numparts
\begin{eqnarray}
\label{eq1}
\ket{e,0} & \rightarrow & \cos{\left(\Omega t /2 \right)} \ket{e,0}
- i \sin{\left(\Omega t /2 \right)} \ket{g,1}, \\[0.1cm]
\label{eq2}
\ket{g,1} & \rightarrow & \cos{\left(\Omega t /2 \right)} \ket{g,1}
- i \sin{\left(\Omega t /2 \right)} \ket{e,0} \, .
\end{eqnarray}
\endnumparts
In the literature, this time evolution of the atom-cavity states is known
also as Rabi rotation where $t$ designates the effective atom-cavity
interaction time in the laboratory and $(\Omega \cdot t)$ the angle of
rotation. Note that neither the state $\ket{e,1}$ nor $\ket{g,0}$ appears
in the time evolution (\ref{eq1})-(\ref{eq2}), in line with our physical
perception that the `photon' of the $e \leftrightarrow g$ transition is
`stored' either by the atom \textit{or} the cavity, but cannot occur twice
in the system.

In contrast to single-mode cavities, a bimodal cavity possesses the
feature of two nearly degenerate light modes of (usually) orthogonal
polarization. Since the frequency of the cavity modes are fixed
geometrically by the design of the cavity, its only the atomic
frequency that can be tuned by means of external fields in order
that the Rydberg atom interacts resonantly with either the first
\textit{or} the second cavity field mode. In practise, this detuning
is done by applying a well adjusted time-varying electric field
across the gap between the cavity mirrors \cite{prl79} so that the
desired (Stark) shift of the atomic transition frequency
$\omega_0(t)$ is achieved. By proper adjustment of $\omega_0(t)$,
one can induce the desired Rabi rotation of the atomic state
interacting either with regard to the first or the second cavity mode
\cite{pra64}. In fact, the development of bimodal cavities has been found
important not only for the manipulation of complex quantum states
but provides one also with an additional `photonic qubit' in the
framework of quantum information that may interact independently with
the Rydberg atoms (`atomic qubits'). Below, we shall denote the
two cavity modes by $M_{1}$ and $M_{2}$ and suppose that they are
associated with the frequencies $\omega_1$ and $\omega_2$, such that
$\omega_1 - \omega_2 \equiv \delta > 0$. Since, moreover, the
frequencies of the two field modes are fixed, only the atomic
$e\leftrightarrow g$ transition frequency can be changed and, as mentioned
before, we shall briefly refer to the (de-)tuning of the atomic
frequency with regard to the cavity modes as \textit{atom-cavity
detuning}.

For a resonant interaction of the Rydberg atoms with both cavity
modes, one need to be able to switch the atomic frequency from one
to the other mode. Let $\Delta(t)$ denote the (time-dependent)
atom-cavity detuning and let us assume that the atom is in resonance
with the cavity mode $M_{1}$ for $\Delta(t) \equiv \omega_0(t) -
\omega_1 = 0$. It is then in resonance with $M_{2}$ for $\Delta(t) =
-\delta$. For a sufficiently large frequency shift $\delta$ of the
two cavity modes, moreover, the atom can be in resonance only with
one of the modes and, hence, the time evolution of the (interacting)
atom-cavity system $A - M_{1} - M_{2}$ can be `divided' into
separate time evolutions owing to the $A - M_{1}$ or $A - M_{2}$
resonant interactions \cite{pra64}. In practise, however, neither
mode of the cavity can be \textit{frozen out} completely because of
the rather small shift $\delta$ ($\approx 2 \, \Omega$ in the
experiments in Refs.\ \cite{haroche, pra64, haroche1}), the decoherence
in the photon field \cite{pra70} as well as imperfections of the cavity
mirrors, i.e.\ the local roughness and deviations from the spherical
geometry. For any theoretical treatment of the decoherence and,
especially, the imperfections of the cavity, we must proceed
\textit{beyond} the Jaynes-Cummings model (\ref{ham}) by allowing
the atom to couple with both cavity modes at the same time
\cite{pra76, pla339}. Such a coupling then supports also the coupling
between the cavity modes and is therefore referred to as
communication channel in the literature \cite{pra70}, cf.\
Section~2.3. It is the goal of this work to demonstrate how a
realistic atom-cavity interaction (coupling) can be analyzed
quantitatively and can improve the theoretical predictions on the
various probabilities, when compared with experiment \cite{pra64}.

\subsection{The step-wise model for detuning the atom-cavity interaction}

\begin{figure}
\begin{center}
\includegraphics[width=1\textwidth]{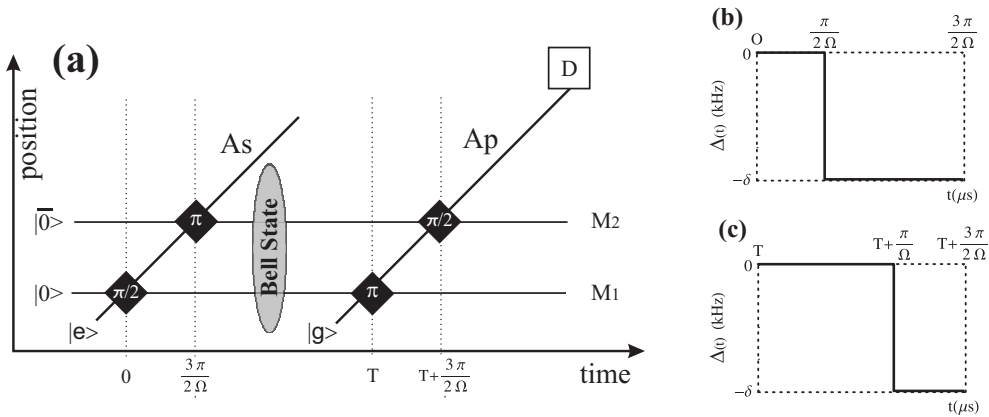}
\\[0.1cm]
\caption{(a) Temporal sequence for producing and observing an entangled
state of the two cavity modes. The pictograms are described in the text.
The grey ellipse area indicates the place where the entangled state between
the cavity modes is obtained. Plots (b) and (c) display the detailed shape
of the atom-cavity detuning $\Delta(t) = \omega_{0}(t) - \omega_{1}$
during the time periods $\left[ 0, \frac{3 \pi}{2 \Omega} \right]$ and
$\left[ T, T+\frac{3 \pi}{2 \Omega} \right]$, respectively.}
\label{fig:1}
\end{center}
\end{figure}

With this short reminder on the development of bimodal cavities and the
Jaynes-Cummings model for the atom-cavity interaction, we are now prepared
to discuss all necessary steps for obtaining the coherent superposition
of the cavity mode state as reported recently in the experiments of
Rauschenbeutel \etal{} \cite{pra64}. In this reference, an \textit{idealized}
time evolution of the atom-cavity system was considered, and it was shown
how the two cavity modes can be brought into the maximally entangled
Bell state
\begin{equation}\label{wave}
\hspace{2cm} \ket{\Psi} = \frac{1}{\sqrt{2}}
    \left( e^{i \psi} \ket{0,\bar{1}} +
    \ket{1, \bar{0}} \right) \, ,
\end{equation}
where $\ket{0},\: \ket{1}$ denote the Fock states of the cavity mode $M_1$,
and $\ket{\bar{0}},\: \ket{\bar{1}}$ the Fock states of mode $M_2$,
respectively.

In a step-wise model of the atom-cavity interaction, the coherent
superposition (\ref{wave}) can be produced as follows. In the
derivations below, we shall often apply the matrix representation of
the time evolution operators in order to discuss the individual
steps. The use of the matrix representation later facilitates also
the application of our two models concerning a more realistic
atom-cavity interaction. Suppose the cavity is initially `empty',
i.e.\ in the state $\ket{0, \bar{0}} \equiv \ket{0} \times
\ket{\bar{0}}$, and is crossed by the (so-called) source atom $A_s$
prepared in the excited state $\ket{e}$. Being within the cavity,
this atom is first tuned in resonance with the cavity mode $M_1$
($\Delta = 0$) for a Rabi rotation $\Omega \, t_1 = \pi/2$, and
followed by a second rotation $\Omega \, t_2 = \pi$, now being in
resonance with the second mode $M_2$ ($\Delta = -\delta$). In the
step-wise model of the atom-cavity interaction, it is essential that
the detuning from mode $M_1$ to $M_2$ can be achieved
instantaneously, i.e.\ within a period that is completely negligible
to the time of interaction with the individual modes. This idealized
sequence of atom-cavity interactions is shown in
Figure~\ref{fig:1}(a) where the Rabi rotations are denoted by the
black diamonds, containing the angle of rotation, and where the
spatio-temporal evolution can be seen for both, the cavity modes as
well as the atoms. Figure \ref{fig:1}(b) displays the step-wise
change in the atom-cavity detuning from the mode $M_1$ to $M_2$.
After the time $\frac{3 \pi}{2 \Omega}$, the source atom has passed
the cavity (and turns out to be de-coupled from the state of the
cavity modes).

To understand the time evolution of the atom-cavity system as a whole, we can
consider successively the resonant interaction of the atom with the two cavity
modes $M_{1}$ and $M_{2}$. From Figure \ref{fig:1}(b), we see that the
(time-dependent) Hamiltonian can be written in the form
\begin{equation} \label{ham1}
\fl \hspace{1.5cm} H_{s}(t) = \theta_{1}(t) H_{1} + \theta_{2}(t)
H_{2} = \cases{ H_1 ( \omega_{0} = \omega_{1} ) , & $0 \leq t \leq
\frac{\pi}{2 \Omega}$
\\ H_2 ( \omega_{0} = \omega_{2} ), &
$\frac{\pi}{2 \Omega} < t \leq \frac{3\pi}{2 \Omega}$ \\},
\end{equation}
with the two `step functions'
\begin{eqnarray}\label{ham2a}
\quad \theta_{1}(t) & = & 1 - \theta \left( t - \frac{\pi}{2\Omega}
\right), \quad \theta_{2}(t) \; = \;   \theta \left( t -
\frac{\pi}{2 \Omega} \right) \, , \nonumber
\end{eqnarray}
and where ($\hbar \equiv 1;\; \mu =1,2$)
\begin{eqnarray}\label{ham2}
H_{\mu} & = & \omega_{0} S_{z}
+ \frac{\Omega}{2} \left( S_{+} a_{\mu} + a_{\mu}^{+} S_{-} \right)
+ \omega_{\mu} \left(a_{\mu}^{+} a_{\mu} + \frac{1}{2} \right)
\end{eqnarray}
refers to the Jaynes-Cummings Hamiltonian as discussed above.
Following the text by Puri \cite{puri}, here we introduce the
operators
\begin{equation}
\fl \hspace{5cm} N_{\mu} = a_{\mu}^{+} a_{\mu} + S_{z} + \frac{1}{2}
\end{equation}
such that $[H_{\mu}, N_{\mu}] = 0$. Especially the last relation enables
us in the case that the atom is in resonance with the field mode
$\left( \omega_{0} = \omega_{\mu} \right)$ to cast the Hamiltonian
(\ref{ham2}) into the equivalent but more convenient form
\begin{equation}
\label{ham3a} \hspace{1.5cm} H_{\mu} = \omega_{\mu} N_{\mu} +
\frac{\Omega}{2} \left( a_{\mu}^{+} S_{-} + S_{+} a_{\mu} \right) \,
.
\end{equation}
Of course, the Hilbert space associated to the Hamiltonian
(\ref{ham1}) is given by the product space of the three
(sub)systems: $A_{s}\; (\ket{g}, \ket{e} )$, $M_{1} \; ( \ket{0},
\ket{1} )$, and $M_{2}\; (\ket{\bar{0}}, \ket{\bar{1}} )$. For the
sake of convenience, let us introduce an explicit notation for the
$2^3$ basis states
\numparts
\begin{eqnarray}
\fl \qquad  \ket{\mathbf{V}_1} = \ket{e,0, \bar{0}}, \quad
  \ket{\mathbf{V}_2} = \ket{g, 1,\bar{1}}, \quad
  \ket{\mathbf{V}_3} = \ket{e, 1, \bar{1}}, \quad
  \ket{\mathbf{V}_4} = \ket{g, 0, \bar{0}}, && \label{basis1} \\
\fl \qquad  \ket{\mathbf{V}_5} = \ket{e, 1, \bar{0}}, \quad
  \ket{\mathbf{V}_6} = \ket{g, 0, \bar{1}}, \quad
  \ket{\mathbf{V}_7} = \ket{e, 0, \bar{1}}, \quad
  \ket{\mathbf{V}_8} = \ket{g, 1, \bar{0}}, && \label{basis2}
\end{eqnarray}
\endnumparts
which naturally fulfill the relations $(i,j = 1, \ldots , 8)$
\begin{eqnarray}
\braket{\mathbf{V}_{i}}{\mathbf{V}_{j}} & = & \delta_{ij}
\qquad {\rm and} \qquad
\sum_{i} \ket{\mathbf{V}_{i}} \bra{\mathbf{V}_{i}} = I \, .
\end{eqnarray}
Using these basis states, the wave function (of the atom-cavity system)
can be always written in the form
\begin{equation} \label{wave0}
\fl \hspace{2cm} \ket{\Psi(t)} \:=\: \sum_{i} c_{i}(t) \ket{\mathbf{V}_i}
\qquad {\rm with} \qquad
c_{i}(t) \:=\: \sum_{j} U_{ij}(t) \, c_{j}(0) \, , \nonumber
\end{equation}
and where $U_{ij}(t) \,=\, \bra{\mathbf{V}_i} U(t)
\ket{\mathbf{V}_j}$ refers to the matrix representation of the time
evolution operator $U(t) = \exp \left( \frac{1}{i} \int_{0}^{t} H(z)
\, dz \right)$. By this definition, the time-evolution matrix
$U^{(s)}_{ij}(t)$ for the atom-cavity state at $t=\frac{3 \pi}{2
\Omega}$ as given by the time-dependent Hamiltonian (\ref{ham1}) is
factorized
\begin{eqnarray}\label{evol1}
U^{(s)}_{ij} \left( \frac{3 \pi}{2 \Omega} \right) & = & \sum_{k}
U^{(2)}_{ik} \left( \frac{\pi}{\Omega} \right) \,
         U^{(1)}_{kj} \left( \frac{\pi}{2 \Omega} \right) \, ,
\end{eqnarray}
where the operator $U^{(1)}(t)$ refers to the evolution due to the
$A_{s}-M_{1}$ interaction and $U^{(2)}(t)$ to that due to
$A_{s}-M_{2}$. Since, as mentioned above, the initial atom-cavity
state is $\ket{\mathbf{V}_1} = \ket{e, 0, \bar{0}}$, we find
$c_{i}(0) = \delta_{i1}$, and the factorization in (\ref{evol1}) is
equivalent to the requirement
\begin{equation}\label{req}
\fl \hspace{3.5cm} \left[ \int_{0}^{t} \theta_{1}(z) \, H_{1} \, dz,
\;
       \int_{0}^{t} \theta_{2}(z) \, H_{2} \, dz \right] \;=\; 0 \, ,
\end{equation}
in accordance with the Hamiltonian (\ref{ham1}).

For the separate steps of the resonant atom-cavity interaction
described by the Hamiltonian (\ref{ham3a}), the Schr\"{o}dinger
equation is given by solutions (\ref{eq1})-(\ref{eq2}). Omitting the
details of the derivation, for which we refer the reader to the
literature \cite{puri}, the matrix representation of the operators
$U^{\,(1)}(t)$ and $U^{\,(2)}(t)$ in the basis
(\ref{basis1})-(\ref{basis2}) is given by
\begin{equation}\label{emat1}
U^{\,(1)}_{ij}(t) = \left(
\begin{array}{cccccccc}
x(t)       & 0     & 0     & 0     & 0     & 0     & 0     & y(t) \\
0          & \bar{x}(t) & 0& 0     & 0     & 0     & \bar{y}(t) & 0 \\
0          & 0     & 1     & 0     & 0     & 0     & 0     & 0 \\
0          & 0     & 0     & 1     & 0     & 0     & 0     & 0 \\
0          & 0     & 0     & 0     & 1     & 0     & 0     & 0 \\
0          & 0     & 0     & 0     & 0     & 1     & 0     & 0 \\
0          & y(t)  & 0     & 0     & 0     & 0     & x(t)  & 0 \\
\bar{y}(t) & 0     & 0     & 0     & 0     & 0     & 0     &
\bar{x}(t)
\end{array}
\right)
\end{equation}
and
\begin{equation}\label{emat2}
U^{(2)}_{ij}(t) = \left(
\begin{array}{cccccccc}
x'(t) & 0     & 0     & 0     & 0     & \bar{y}'(t) & 0    & 0 \\
0     & \bar{x}'(t)   & 0     & 0     & y'(t)  & 0  & 0    & 0 \\
0     & 0     & 1     & 0     & 0     & 0      & 0  & 0 \\
0     & 0     & 0     & 1     & 0     & 0      & 0  & 0 \\
0     & \bar{y}'(t)   & 0     & 0     & x'(t)  & 0  & 0    & 0 \\
y'(t) & 0     & 0     & 0     & 0     & \bar{x}'(t) & 0    & 0 \\
0     & 0     & 0     & 0     & 0     & 0      & 1  & 0 \\
0     & 0     & 0     & 0     & 0     & 0      & 0  & 1
\end{array}
\right) \, ,
\end{equation}
and where
\begin{equation}\label{sols0}
x(t) = \bar{x}(t) = \cos \left( \frac{\Omega t}{2} \right), \qquad
y(t) = \bar{y}(t) = -i \sin \left( \frac{\Omega t}{2} \right) \, .
\end{equation}
In this notation, the prime refers to the additional phase factor
$\Box^{\prime}(t) = e^{i \delta t} \cdot \Box(t)$ that arises from the energy
difference $\hbar \delta$ of the two cavity modes and is accumulated during
the $(A - M_2)$ Rabi rotation \cite{pra64}. Substituting the expressions
(\ref{evol1}), (\ref{emat1}) and (\ref{emat2}) into the wave function
(\ref{wave0}) and by making use of the initial condition
$c_{i}(0) = \delta_{i1}$, we then obtain at the time
$t = \frac{3 \pi}{2\Omega}$ two non-zero coefficients
\begin{equation}
c_{6} \left( \frac{3 \pi}{2 \Omega} \right) = - \frac{i}{\sqrt{2}}
e^{\frac{i \delta \pi}{\Omega}}, \qquad c_{8} \left( \frac{3 \pi}{2
\Omega} \right) = - \frac{i}{\sqrt{2}},
\end{equation}
and, hence, the total wave function (up to an irrelevant phase
factor) is
\begin{equation}\label{wave1}
\ket{\Psi \left( \frac{3 \pi}{2 \Omega} \right)} =
\frac{1}{\sqrt{2}}
    \left( e^{\frac{i \delta \pi}{\Omega}}
           \ket{0,\bar{1}} + \ket{1, \bar{0}} \right) \ket{g} \, .
\end{equation}
This wave function is equivalent to those derived in \cite{pra64}
and is the same as displayed above (\ref{wave}). Apparently, the two
cavity modes form now are maximally entangled state, while (the
state of) the source atom is factorized out.

Unfortunately, the cavity state is inaccessible for direct
measurements. In order to `prove' the coherent superposition
(\ref{wave1}) of the two cavity modes, another (probe) atom $A_p$
has to be sent through the cavity after a time delay $\left( T -
\frac{3 \pi}{2 \Omega} \right)$. In the experiments in \cite{pra64},
this time delay was chosen between about $0\, .. \, 710 \:\mics$.
The purpose of the probe atom is to `read off' the state
of the cavity modes and to copy this information upon its own state.
Similar as the source atom, the probe atom $A_p$ interacts with the
cavity  during the time interval $\left[ T, T+\frac{3 \pi}{2 \Omega}
\right]$ being in resonance with the cavity modes $M_1$ and $M_2$ as
shown in Figure {\ref{fig:1}}(c). Using the arguments from above,
the whole time evolution of the atom-cavity system from zero up to
the time $\left( T + \frac{3 \pi}{2 \Omega} \right)$ is given by
applying the evolution matrix
\begin{equation}\label{evol2}
\fl \qquad U_{ij} \left( T + \frac{3 \pi}{2 \Omega} \right) =
\sum_{k,m,l} \, U^{(5)}_{ik} \left( \frac{\pi}{2 \Omega} \right)
                U^{(4)}_{km} \left( \frac{\pi}{ \Omega}  \right)
\, U^{(3)}_{ml} \left( T - \frac{3 \pi}{2 \Omega} \right)
        U^{(s)}_{lj} \left( \frac{3 \pi}{2 \Omega} \right) \,.
\end{equation}
In this matrix, $U^{(s)}_{lj}$ is given by (\ref{evol1}) and refers
to the interaction of the cavity with the source atom, while the
other matrices $U^{(3)}_{ml},\; U^{(4)}_{km}$ and $U^{(5)}_{ik}$
describe the three subsequent steps: the \textit{free} time
evolution during the time interval $\left[\frac{3 \pi}{2 \Omega}, T
\right]$
\begin{equation}\label{emat4}
\fl \hspace{4cm} U^{(3)}_{ij}(t) = {\rm diag} \{ 1, 1, 1, 1, 1, e^{i
\delta t}, 1, 1 \} \, ,
\end{equation}
which is diagonal in the basis (\ref{basis1})-(\ref{basis2}) and
where a relative phase shift $e^{i \delta t}$ is accumulated due to
the energy difference $\hbar \delta$ of the two cavity modes.
Afterwards, during the interval $\left[T, T + \frac{3 \pi}{2 \Omega}
\right]$, the further evolution of the atom-cavity system is driven
by the Hamiltonian [cf.\ Figure~\ref{fig:1}(c)]
\begin{equation}
\fl \hspace{2.0cm} H_{p}(t) =
    \theta_{3}(t) H_{1} + \theta_{4}(t) H_{2} =
    \cases{H_1, & $T \leq t \leq T + \frac{\pi}{\Omega}$ \\
           H_2, & $T + \frac{\pi}{\Omega} < t \leq
               T + \frac{3\pi}{2\Omega}$  \\ } \, .
\end{equation}
Since the atom-cavity interaction is as well described by the
Jaynes-Cummings Hamiltonian (\ref{ham3a}), the matrices
$U^{(4)}_{ij}(t)$ and $U^{(5)}_{ij}(t)$ coincides with the matrices
(\ref{emat1}) and (\ref{emat2}), except that in $U^{(4)}_{ij}(t)$
the relative phase $e^{i\delta t}$ occurs. Including this phase
factor, $U^{(4)}_{ij}(t)$ then becomes
\begin{equation}\label{emat3}
U^{(4)}_{ij}(t) = \left(
\begin{array}{cccccccc}
x(t)  & 0     & 0     & 0     & 0     & 0     & 0     & y(t) \\
0     & \bar{x}(t)  & 0     & 0     & 0     & 0     & \bar{y}(t)   & 0 \\
0     & 0     & 1     & 0     & 0     & 0     & 0     & 0 \\
0     & 0     & 0     & 1     & 0     & 0     & 0     & 0 \\
0     & 0     & 0     & 0     & 1     & 0     & 0     & 0 \\
0     & 0     & 0     & 0     & 0     & e^{i \delta t}     & 0     & 0 \\
0     & y(t)  & 0     & 0     & 0     & 0     & x(t)  & 0 \\
\bar{y}(t)  & 0     & 0     & 0     & 0     & 0     & 0     &
\bar{x}(t)
\end{array}
\right) \, .
\end{equation}

Having the complete time evolution (\ref{evol2}), we can easily
determine the probability $P(T)$ to find the probe atom in the
exited state $\ket{e}$, after it has crossed the cavity and hit the
detector D.  This is the (final-state) probability that has been
measured during the experiment \cite{pra64} and that provides us
with the information about the superposition of the cavity mode
states (\ref{wave1}). In practise, of course, the atom-cavity
interaction is not \textit{ideal} and has to be replaced by some
realistic model for the atom-cavity interaction. For the idealized
sequence of interaction, however, the (complete) time evolution
owing to the matrix (\ref{evol2}) at the time $ t = \frac{3 \pi}{2}
+ T$ implies only the two non-zero coefficients
\begin{eqnarray}
c_{1} \left( \frac{3 \pi}{2} + T \right) & = & - \frac{1}{2}
e^{\frac{i \delta \pi}{2\Omega}}
  \left(  1 + e^{\frac{i \delta \pi}{2\Omega}} e^{i \delta T}
  \right),
\nonumber \\[0.1cm]
c_{6} \left( \frac{3 \pi}{2} + T \right) & = &
  \frac{i}{2} e^{\frac{i \delta \pi}{2\Omega}}
  \left( 1 - e^{\frac{i \delta \pi}{2\Omega}} e^{i\delta T} \right);
\nonumber
\end{eqnarray}
and, hence, the final-state probability to find $A_{p}$ in the excited
state is simply
\begin{equation}\label{prob0}
P(T) = \left| c_{1} \left( \frac{3 \pi}{2} + T \right) \right|^{2}
     = \frac{1 + \cos \left( \omega \, T + \phi \right)}{2} \, .
\end{equation}
Obviously, this probability oscillates between zero and one with the
frequency $\omega \equiv \delta$  and with the phase (shift) $\phi
\equiv \pi \delta / 2 \Omega$. In the experiments by Rauschenbeutel
\etal{} \cite{pra64} and Raimond \etal{} \cite{haroche}, the shift
between the cavity mode frequencies takes the value $\delta /2 \pi =
128.3$ kHz, while the atom-cavity coupling constant is $\Omega /
2\pi = 47$ kHz.

\subsection{Changing the atom-cavity detuning in a smooth way}

\begin{figure}
\begin{center}
\includegraphics[width=0.65\textwidth]{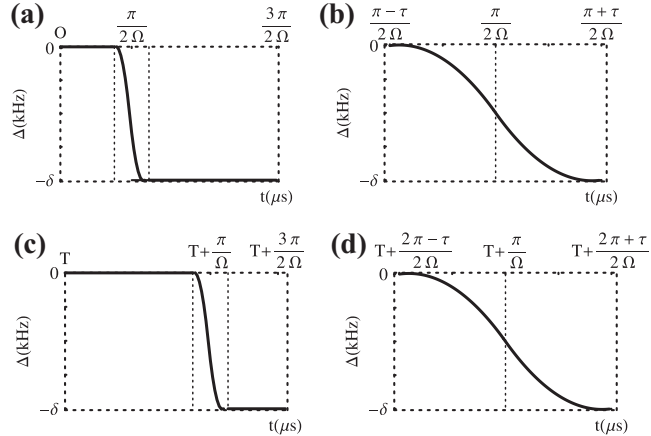}\\
\caption{Temporal sequence of the atom-cavity detuning if the interaction of
the atom is changed smoothly from one to the other cavity mode. Plot (a)
display the atom-cavity detuning $\Delta(t) = \omega_0(t) - \omega_1$ during
the interval $\left[ 0, \frac{3\pi}{2 \Omega} \right]$ where the transition
frequency of the source atom is detuned from the cavity mode $M_1$ to
the mode $M_2$; this `smooth change' is inflated in plot (b) for the
interval $\left[\frac{\pi-\tau}{2\Omega}, \frac{\pi-\tau}{2\Omega} \right]$.
The plots (c) and (d) display the same but for the atom-cavity detuning
$\Delta(t) = \omega_0(t) - \omega_1$ during the time interval
$\left[ T, T+\frac{3 \pi}{2 \Omega} \right]$, i.e.\ when the cavity
interacts with the probe atom.}
\label{fig:2}
\end{center}
\end{figure}

In the previous Subsection, it was shown how the (two) cavity modes
become entangled with each other by detuning the atom-cavity
interaction in a proper way. For a step-wise detuning $\Delta(t)$ as
displayed in figures \ref{fig:1}(b) and (c), this leads to the
maximally entangled state (\ref{wave1}) of the two cavity modes
after the source atom has passed through the cavity, leaving the
atom factorized out in a product state. This `sudden' change in the
detuning $\Delta(t)$ is however not physically feasible, and further
analysis is required to understand how the evolution of cavity
states is affected by a more realistic time-dependence of the
detuning. In this sections, therefore, we shall consider a smooth
`switch' from the $A_{s/p}-M_{1}$ to the $A_{s/p}-M_{2}$ resonant
interaction within a \textit{finite} time period, which we will
denote by $\tau$. For instance, according to the experimental setup
\cite{pra64} this period is of the length $\tau / \Omega \leq 1
\mics$, which corresponds to $\sim$ 7~\%{} of the overall $\frac{3
\pi}{2 \Omega}$ interaction time of the atom with the cavity, or
about an angle $\frac{\pi}{10}$ in units of Rabi rotations, which is
not anymore negligible (as we considered in the previous model).
Below, we shall assume a smooth behavior for this finite switch as
displayed in figures \ref{fig:2}(a) and (c), where the cavity is
non-resonant to the source atom during the interval $\left[
\frac{\pi - \tau}{2\Omega}, \frac{\pi + \tau}{2 \Omega}\right]$ and
with the probe atom during $\left[ T + \frac{2\pi - \tau}{2 \Omega},
        T + \frac{2\pi + \tau}{2 \Omega} \right]$, respectively.
This smooth switch of the atomic resonance from one to the other
cavity mode can be described mathematically by the Hamiltonian
\begin{equation}
\label{ham3}
H^{\tau}_{s}(t) =
\cases{H_{1},    & $0 \leq t \leq \frac{\pi - \tau}{2 \Omega}$ \\
       H_{-}(t), & $\frac{\pi - \tau}{2 \Omega} < t \leq
                    \frac{\pi}{2\Omega}$                       \\
       H_{+}(t), & $\frac{\pi}{2 \Omega} < t \leq
                    \frac{\pi + \tau}{2\Omega}$                \\
       H_{2},    & $\frac{\pi + \tau}{2 \Omega} < t \leq
                    \frac{3\pi}{2 \Omega}$                     \\ } \, ,
\end{equation}
where $H_{1}$ and $H_{2}$ are the resonant Jaynes-Cummings
Hamiltonian (\ref{ham3a}) from above, and where $H_{-}(t)$ and $H_{+}(t)$
are given by
\begin{eqnarray}
\label{ham6a}
&& \quad H_{-}(t) = S_{z} \Delta(t) + \omega_{1} N_{1}
   + \frac{\Omega}{2}
   \left( a_{1}^{+} S_{-} + S_{+} a_{1} \right), \\[0.1cm]
\label{ham6b}
&& H_{+}(t) = S_{z} \left[ \Delta(t) + \delta \right]
   + \omega_{2} N_{2} + \frac{\Omega}{2}
   \left( a_{2}^{+} S_{-} + S_{+} a_{2} \right)  \, .
\end{eqnarray}
Here, the superscript `$\tau$' in (\ref{ham3}) is introduced in
order to distinguish the Hamiltonian (\ref{ham3}) from that of the
step-wise detuning (\ref{ham1}). Note that, for the present smooth
change of the atom-cavity detuning, $H_1$ and $H_{-}(t)$
\textit{together} describe the (time) evolution of the $A_s - M_1$
interaction up to the time $\frac{\pi}{2\Omega}$. That is the
resonant part of the atom-cavity interaction due to $H_1$ is
shortened by the time interval $\frac{\tau}{2 \Omega}$, while the
remaining time up to $\frac{\pi}{2\Omega}$ follows the
\textit{non-resonant} Jaynes-Cummings Hamiltonian $H_{-}(t)$ in line
with figure \ref{fig:2}(a)-(b). Similarly, the $A_s - M_2$
interaction is first non-resonant and modeled by the Hamiltonian
$H_{+}(t)$ for the interval $[ \frac{\pi}{2\Omega},
\frac{\pi+\tau}{2\Omega}]$, and becomes later resonant with the
cavity mode $M_2$, as seen on figure \ref{fig:2}(c)-(d). Obviously,
the Hamiltonian (\ref{ham3}) can be understood also as an extension
of the `step-wise' model from Section 2.1 by introducing an
additional short period for the non-resonant atom-cavity
interaction. For the Hamiltonian (\ref{ham3}), therefore, the time
evolution matrix still factorizes and can be expressed at $t =
\frac{3 \pi}{2 \Omega}$ as
\begin{equation}\label{evol3}
\fl \quad U^{(s)}_{ij} \left( \frac{3 \pi}{2 \Omega} \right) =
\sum_{k,l,m} \, U^{(2)}_{ik} \left( \frac{\pi}{\Omega}
                - \frac{\tau}{2\Omega} \right)
        U^{(+)}_{kl} \left( \frac{\tau}{2\Omega} \right) \,
        U^{(-)}_{lm} \left( \frac{\tau}{2\Omega} \right)
                U^{(1)}_{mj} \left( \frac{\pi - \tau}{2 \Omega} \right)
\end{equation}
where the matrices $U^{(1)}_{ij}(t)$ and $U^{(2)}_{ij}(t)$ are the
same as in (\ref{emat1}) and (\ref{emat2}), and where the matrix
$U^{(-)}_{ij}(t)$ has still the form (\ref{emat1}) but with $x(t)$,
$\bar{x}(t)$, $y(t)$, and $\bar{y}(t)$ now being solutions of the
differential equations \cite{puri}:
\begin{eqnarray}\label{eqs}
\fl \qquad i \frac{d y(t)}{d t} & = & \frac{\Omega}{2} \bar{x}(t) +
\frac{1}{2} \Delta(t) \, y(t), \qquad  i \frac{d \bar{x}(t)}{d t} =
\frac{\Omega}{2} y(t) - \frac{1}{2} \Delta(t) \, \bar{x}(t),
\label{eqs1} \\ \fl \qquad i \frac{d x(t)}{d t} & = &
\frac{\Omega}{2} \bar{y}(t) + \frac{1}{2} \Delta(t) \, x(t), \qquad
i \frac{d \bar{y}(t)}{d t} = \frac{\Omega}{2} x(t) - \frac{1}{2}
\Delta(t) \, \bar{y}(t) \, . \label{eqs2}
\end{eqnarray}
Analogue, the matrix $U^{(+)}_{ij}(t)$ has the form (\ref{emat2})
but with functions $x(t)$, $\bar{x}(t)$, $y(t)$, and $\bar{y}(t)$
that are obtained from (\ref{eqs}) and (\ref{eqs2}) by replacing
$\Delta(t) \rightarrow \Delta(t) + \delta$.

No exact analytic solutions are known for the relations
(\ref{eqs1})--(\ref{eqs2}). Therefore, in order to derive the time
evolution matrix (\ref{evol3}), we need to solve these equations for
$x(t)$, $\bar{x}(t)$, $y(t)$, and $\bar{y}(t)$ numerically up to $t
= \frac{\tau}{2\Omega}$, and then to substitute their values into
the matrices $U^{(\pm)}_{ij}(t)$. As discussed above, the
final-state probability to find the probe atom in the state
$\ket{e}$ is obtained by performing a similar procedure for the
$A_{p}-M_{1}-M_{2}$ interaction during the time interval $\left[ T,
T+\frac{3\pi}{2 \Omega} \right]$. During the time $\left[\frac{3
\pi}{2 \Omega}, T \right]$, of course, the `free' evolution of the
cavity state still follows the trivial matrix (\ref{emat4}). Putting
all these pieces together, hence, the complete time evolution of the
wave function for such a smooth change in the atom-cavity
interaction is described by the matrix
\begin{eqnarray}
\label{evol4}
\fl U^{(\tau)}_{ij} \left( T + \frac{3 \pi}{2 \Omega} \right) & = &
\sum_{k,m,l,n,p} \, U^{(5)}_{ik} \left( \frac{\pi - \tau}{2 \Omega}
\right) U^{(+)}_{km} \left( \frac{\tau}{2\Omega} \right) \,
U^{(-)}_{ml} \left( \frac{\tau}{2\Omega} \right) \, U^{(4)}_{ln}
\left( \frac{\pi}{ \Omega} - \frac{\tau}{2 \Omega} \right)
\nonumber \\[0.1cm]
&   & \hspace*{1.75cm} \times \, U^{(3)}_{np} \left( T - \frac{3
\pi}{2 \Omega} \right) \, U^{(s)}_{pj} \left(\frac{3 \pi}{2 \Omega}
\right) \, ,
\end{eqnarray}
and where the matrices $U^{(3)}_{ij}(t)$, $U^{(4)}_{ij}(t)$, $U^{(5)}_{ij}(t)$,
and $U^{(s)}_{ij}(t)$ coincide with matrices (\ref{emat4}), (\ref{emat3}),
(\ref{emat2}), and (\ref{evol3}), respectively. For the initial condition
$c_{i}(0) = \delta_{i1}$, the final-state probability in model is again
\begin{equation}
\label{prob1}
\fl \hspace{3.5cm} P_{\tau}(T) =
|\braket{\mathbf{V}_1}{\Psi_{\tau}(T)}|^{2} \; = \; \left|
c_{1}\left( T + \frac{3 \pi}{2 \Omega} \right) \right|^{2}
\end{equation}
which, however, cannot be given in a closed form because of the need
for solving the functions $x(t)$, $\bar{x}(t)$, $y(t)$, and
$\bar{y}(t)$ numerically in the various matrices. Like the cosine
function (\ref{prob0}) from the step-wise model, this probability
still oscillates between zero and one with the frequency
$\omega_{\tau}$ and phase $\phi_{\tau}$. In the Section 3, we
display and discuss how the frequency and phase (shift) behave as
function of the `switching' parameter $\tau$, i.e.\ the time for
which the cavity is non-resonant to the atom.

\subsection{Communication channel model: Allowing the mutual
            interaction of the cavity modes}

\begin{figure}
\begin{center}
\includegraphics[width=0.65\textwidth]{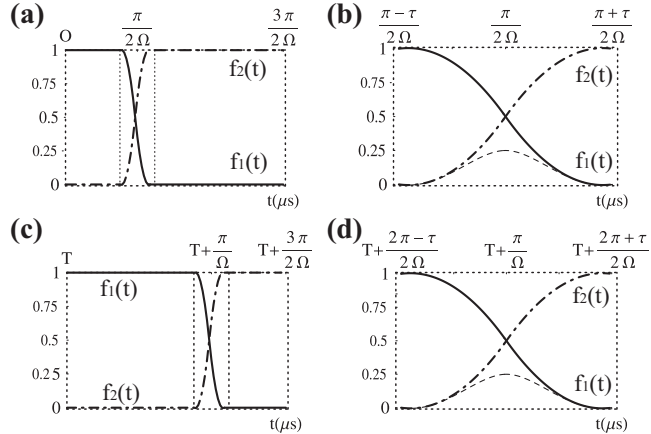}\\
\caption{Temporal shape of the dimensionless functions $f_{1}(t)$
and $f_{2}(t)$ that describe the contribution of the $A_{s/p} - M_1$
interaction (solid line) and the $A_{s/p} - M_2$ interaction
(dot-dashed line) to the total Hamiltonian of the atom-cavity
interaction. Plots (b) and (d) display also temporal shape of the
product $f_{1}(t) \cdot f_{2}(t)$ (dashed line) that describes the
contribution of `communication channel' to the total Hamiltonian of
the atom-cavity interaction.}
\label{fig:3}
\end{center}
\end{figure}

Up to the present, we have always assumed that, at a given time $t$,
the atom interacts either with the cavity mode $M_{1}$ \textit{or}
the mode $M_{2}$, implying a sudden change in the interaction from
one cavity mode to the other. In our notation above, this means that
the overall atom-cavity interaction is understood as a set of
\textit{pure} $A_{s/p} - M_{1}$ or $A_{s/p} - M_{2}$ (non)resonant
interactions. However notice that at the time $t = \frac{\pi}{2
\Omega}$ in figure \ref{fig:2}(b), for instance, the atomic
transition frequency is equally far from the (nearly degenerate)
frequencies of both cavity modes and, hence, we expect the atom to
`feel' the (non-resonant) contribution of $M_1$ and $M_2$
simultaneously. As additional sources for such a simultaneous
interaction also serve the imperfections of the mirrors, stray
fields, decoherence, etc. Finally, this simultaneous interaction
also implies a mutual interaction among the cavity modes to which
one therefore refers as a communication channel \cite{pra70, puri}.

To describe this `communication' between the two cavity modes, we
need to proceed \textit{beyond} the Jaynes-Cummings model in order
to allow the atom to couple with both cavity modes at the same time.
In addition, in order to preserve the `correspondence' with the
previous two models, this simultaneous interaction must happen only
inside the time intervals $\left[ \frac{\pi - \tau}{2\Omega},
\frac{\pi + \tau}{2 \Omega}\right]$ and $\left[ T + \frac{2\pi -
\tau}{2 \Omega}, T + \frac{2\pi + \tau}{2 \Omega} \right]$, given by
the same `switching' parameter $\tau$ as in Section 2.2. Below, we shall model
the total (source) atom-cavity interaction within the interval
$\left[0, \frac{3 \pi}{2 \Omega} \right]$ by means of the
time-dependent Hamiltonian
\begin{equation}
\label{ham5} \widetilde{H}^{\tau}_{s}(t) =
\cases{ H_{1},         & $0 \leq t \leq \frac{\pi - \tau}{2 \Omega}$ \\
        H_{\times}(t), & $\frac{\pi - \tau}{2 \Omega} < t \leq
                          \frac{\pi + \tau}{2 \Omega}$               \\
        H_{2},         & $\frac{\pi + \tau}{2 \Omega} < t \leq
                      \frac{3\pi}{2\Omega}$ \\ }
\end{equation}
with
\begin{eqnarray}
&& H_{\times}(t) = f_{1}(t) H_{-}(t) + f_{2}(t) H_{+}(t)
              + f_{1}(t) f_{2}(t) H_{I}\, ,
\\
\label{inter} && \hspace{2cm} H_{I} = \lambda \left( a^{+}_{1} a_{2}
+ a^{+}_{2} a_{1} \right) \, ,
\end{eqnarray}
and where $H_{\mu}$ and $H_{\pm}(t)$ denote the Hamiltonians
(\ref{ham3a}) and (\ref{ham6a})-(\ref{ham6b}), respectively. The
temporal behavior of the detuning $\Delta(t)$ used in $H_{\pm}(t)$
is the same as we considered in the previous section and which is
displayed in Figure \ref{fig:2}. The two (dimensionless) functions
$f_1(t)$ and $f_2(t)$ displayed in Figure \ref{fig:3}(a), thus
determine the contribution of the non-resonant Jaynes-Cummings
Hamiltonians $H_{\pm}(t)$ and the communication channel Hamiltonian
$H_{I}$ to the overall time-dependent Hamiltonian (\ref{ham5}). In
this model, again, the resonant $A_s - M_1$ and $A_s - M_2$
interactions are shortened by time interval $\tau / \Omega$ and the
evolution of the (atom-cavity) state is governed by the Hamiltonian
$H_{\times}(t)$ with the non-resonant part $H_{\pm}(t)$ and the
communication channel $H_{I}$. The Hamiltonian $H_{I}$ is the
simplest expression that fulfills the above requirements and
preserves the cavity field energy. The contribution of $H_{I}$ to
the Hamiltonian $H_{\times}(t)$ is given by the time-dependent
dimensionless product $f_1(t) \, f_2(t)$. The shape of this
coupling term is displayed in Figure \ref{fig:3}(b) by a dashed line,
which has its maximal value $1/4$ at $t_{max} = \pi / 2 \Omega$.
Making use of the functions $f_{\mu}(t)$ in the Hamiltonian (\ref{ham5}),
we expect to describe the atom-cavity interaction during the switching
period $\tau / \Omega$ in a realistic fashion. Moreover, below we shall
consider also the coupling strength $\lambda$ of the cavity modes
such that the `effective' coupling $\lambda \, f_1(t) \, f_2(t)$
satisfies the relation
\begin{equation}
\hspace{2cm} \lambda \, f_1(t_{max}) \, f_2(t_{max}) = \Omega.
\end{equation}
The meaning of the last relations can be understood as follows: The
strongest $M_1 - M_2$ effective coupling should be (or, at least,
should not exceed) the composition of the $A_s - M_1$ and $A_s - M_2$
couplings given by the values of $\Omega/2$, respectively. A simple
calculation implies the value of $\lambda$ to be of the order of
$\sim$ 1 MHz. For this value, indeed, a good agreement with the
experimental data of \cite{pra64} has been found within the expected time
$(\tau / \Omega \leq 1 \mics)$, i.e.\ for that time which is needed
to `switch' the atomic frequency between the two cavity
modes (see Section 3). Therefore, the Hamiltonian (\ref{ham5}) exposes
the right `correspondence' property with regard to the models we
have introduced in the previous two Sections. This can be readily seen
if we assume $f_{1}(t) = \theta_{1}(t)$ and $f_{2}(t) = \theta_{2}(t)$;
then, expression (\ref{ham5}) reduces to the Hamiltonian
$H^{\tau}_{s}(t)$ (\ref{ham3}) and, if we additionally impose the
limit $\tau \rightarrow 0$, the Hamiltonian (\ref{ham5}) simplifies to
$H_{s}(t)$ in expression (\ref{ham1}).

The time evolution matrix at $t = \frac{3 \pi}{2 \Omega}$ as
described by the Hamiltonian (\ref{ham5}) is given by
\begin{equation}\label{evol5}
\fl \hspace{2cm} \widetilde{U}^{(s)}_{ij} \left( \frac{3 \pi}{2
\Omega} \right) = \sum_{k,l} U^{(2)}_{ik} \left(\frac{\pi}{\Omega} -
\frac{\tau}{2\Omega} \right) \, U^{(\times)}_{kl} \left(
\frac{\tau}{\Omega} \right) \, U^{(1)}_{lj} \left( \frac{\pi -
\tau}{2 \Omega} \right) \, ,
\end{equation}
where $U^{(1)}_{ij}(t)$ and $U^{(2)}_{ij}(t)$ refer to the expressions
(\ref{emat1}) and (\ref{emat2}), and where the matrix elements of
$U^{(\times)}_{ij}(t)$ must fulfill the Schr\"{o}dinger equation
\begin{equation}
\label{U-Schroedinger} \qquad i \, \frac{d U^{(\times)}_{ij}(t)}{dt}
= \sum_{k} \, \bra{\mathbf{V}_i} H_{\times}(t) \ket{\mathbf{V}_k} \,
U^{(\times)}_{kj}(t) \, .
\end{equation}
This set of differential equations cannot be solved analytically and
need to be integrated numerically during the interval $t =
\frac{\tau}{\Omega}$. Indeed, a similar procedure has to be
performed later also for the $A_{p}-M_{1}-M_{2}$ interaction during
the time $\left[T, T+\frac{3\pi}{2 \Omega} \right]$ as seen from
figures \ref{fig:3}(c),(d) and figures \ref{fig:2}(c),(d). Combining
the pieces together, the complete time evolution of the atom-cavity
interaction therefore becomes
\begin{eqnarray}\label{evol6}
\fl \qquad \widetilde{U}^{(\tau)}_{ij} \left( T + \frac{3 \pi}{2
\Omega} \right) & = & \sum_{k,m,l,n} U^{(5)}_{ik} \left( \frac{\pi -
\tau}{2\Omega} \right) \, U^{(\times)}_{km} \left(
\frac{\tau}{\Omega} \right) \, U^{(4)}_{ml} \left( \frac{\pi}{
\Omega} - \frac{\tau}{2 \Omega}
\right) \nonumber \\[0.1cm]
&   & \hspace*{0.75cm} \times \, U^{(3)}_{ln} \left( T -
\frac{3\pi}{2 \Omega} \right) \, \widetilde{U}^{(s)}_{nj} \left(
\frac{3\pi}{2 \Omega} \right) \, ,
\end{eqnarray}
where the matrices $U^{(3)}_{ij}(t)$, $U^{(4)}_{ij}(t)$, $U^{(5)}_{ij}(t)$,
and $\widetilde{U}^{(s)}_{ij}(t)$ coincide with the matrices (\ref{emat4}),
(\ref{emat3}), (\ref{emat2}), and (\ref{evol5}) respectively. Moreover, the
final-state probability is given for the initial conditions
$c_{i}(0) = \delta_{i1}$ by
\begin{equation}
\label{prob2} \widetilde{P}_{\tau}(T) \:=\:
|\braket{\mathbf{V}_1}{\widetilde{\Psi}_{\tau}(T)}|^{2} \:=\: \left|
c_{1}\left( T + \frac{3 \pi}{2 \Omega} \right) \right|^{2} \, .
\end{equation}
Although no closed expression can be given for this probability (due
to the numerical parts in the time evolution), it still oscillates
between zero and one with the frequency $\widetilde{\omega}_{\tau}$
and the phase $\widetilde{\phi}_{\tau}$ that is different from the
two other models in Sections 2.1 and 2.2. In the next Section, we
shall discuss how these frequency and phase (shifts) behave as
functions of the `switching' parameter $\tau$.

\section{Comparison with experiment and discussion}

\begin{figure}
\begin{center}
\includegraphics[width=0.7\textwidth]{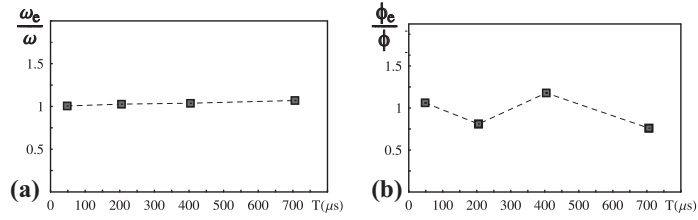} \\
\caption{The relative frequency (a) and phase (b) of the final-state
probability as obtained experimentally and displayed for the four time
intervals $I_{\nu}$ in \cite{pra64}. The values from the experiment are
displayed by dots and are combined by a dashed line in order to guide the
eyes of the reader. See text for further explanations.}
\label{fig:4}
\end{center}
\end{figure}

The expression (\ref{prob0}) of the final-state probability
describes the time evolution of an idealized experiment \cite{pra64}
with a step-wise change in the resonant interaction between the two
cavity modes. This expression neglects of course a number of
additional effects, such as the cavity field relaxation, the
influence of external stray fields, or imperfections due to the
cavity mirrors and cavity geometry. The most significant distortion
of the idealized probability in (\ref{prob0}) is given by the cavity
field relaxation during the interval $\left[ \frac{3 \pi}{2 \Omega},
T \right]$, i.e. during the `free' evolution of the cavity field,
since during the $A_s - M_1 - M_2$ and $A_p - M_1 - M_2$
interactions, the system evolves in the strong coupling (resonant)
regime, and hence, the dissipation of the cavity energy is
negligible. This field relaxation mainly arises from the interaction
of the cavity with the environment and its effect on the final-state
probability (\ref{prob0}) has been investigated recently by
de~Magalhaes and Nemes \cite{pra70}. In this reference, the
environmental degrees of freedom were modeled by a infinite set of
harmonic oscillators whose annihilation (creation) operators $c_b$
($c_b^+$) are coupled linearly to the Hamiltonian of the cavity
field modes as follows
\begin{equation}\label{coupled}
H(t) =
\cases{H_s(t),      & $0 \leq t \leq \frac{3\pi}{2 \Omega}$     \\
       H_{coupled}, & $\frac{3\pi}{2 \Omega} < t < T$           \\
       H_p(t),      & $T \leq t \leq T + \frac{3\pi}{2 \Omega}$ \\}
\end{equation}
with
\begin{equation*}
\fl \quad H_{coupled} = \sum_{\mu=1,2} \omega_{\mu} \, a^+_{\mu}
a_{\mu} + \sum_b^{\infty} \left( \omega_b \, c^+_b c_b + \gamma_{1b}
                       \left[a_1 c^+_b + a^+_1 c_b \right]
             + \gamma_{2b} \left[a_2 c^+_b + a^+_2 c_b \right]
        \right),
\end{equation*}
and where $\omega_b$ describe the frequencies and $\gamma_{\mu b}$
the respective coupling constants. Following \cite{pra70}, the
final-state probability can be cast into the form
\begin{equation}
\label{prob}
\fl \hspace{3.5cm} P(T) =
A(T) \frac{1 + \cos \left( \omega \, T + \phi \right)}{2} + B(T),
\end{equation}
with
\begin{equation}
\label{damp} \fl \qquad A(T) = e^{-(\alpha + \beta) \, \xi}, \quad
B(T) = \frac{e^{-2 \alpha \, \xi} + e^{-2 \beta \, \xi} - 2
e^{-(\alpha + \beta) \, \xi}}{4}, \quad \xi \equiv T - \frac{3
\pi}{2 \Omega},
\end{equation}
and in which $\alpha$, $\beta$ denote the dissipation constants of
the cavity, while the frequency $\omega$ and phase $\phi$ coincides
with those of expression (\ref{prob0}). In fact, the probability
(\ref{prob}) has been found in a good agreement with the
experimental data of \cite{pra64} with regard to the amplitude and
frequency of the oscillations. However, less agreement was obtained
for the phase shift of the final-state probability as given by $\phi
\equiv \pi \delta / 2 \Omega$. Note that, according to (\ref{prob}),
the two arguments of the \textit{cosine}: $\omega$ and $\phi$, are
independent of the dissipation constants $\alpha$ and $\beta$ of the
cavity. Therefore, we can describe the frequency and phase shifts in
the final-state probability separately from the damping of the
signal, as given by functions (\ref{damp}). For the two models
(\ref{prob1}) and (\ref{prob2}), indeed, we will show below that a
finite `switch' (Section 2.2) and the communication between the
cavity modes (Section 2.3) mainly affects the phase of the amplitude
(\ref{prob}), as have been prognosticated in \cite{pra70}.
Therefore, combined with the `damping functions' (\ref{damp}), our
investigations enables us to further improve the agreement with
experiment.

We now compare the predictions for the frequency and phase of the
final-state probability, obtained with the models from Section~2,
with those from experiment \cite{pra64}. In these measurements, the
probability for detecting the probe atom $A_{p}$ in the excited
state $\ket{e}$ was recorded and displayed for the four time
intervals $I_1 = [48; 57] \, \mics$, $I_2 = [200; 207]\, \mics$,
$I_3 = [400, 408]\, \mics$, and $I_4 = [699,706]\, \mics$,
respectively. From the observed probability (cf.\ Figures 2(a)-(d)
in \cite{pra64}), the four `experimental' frequencies and phases
have been extracted by a fit to the analytical function
(\ref{prob}), where the cosine arguments have been considered as
unknown values, separately for each time interval $I_{\nu}$ ($\nu =
1, \ldots, 4$), and are displayed by the dots in figure \ref{fig:4}.
For the sake of convenience, moreover, these experimental data were
normalized to the reference frequency $\omega \equiv \delta = 2 \pi
\times 128.3$ kHz and phase $\phi \equiv \pi \delta / 2 \Omega =
4.29$ of the `idealized' probability (\ref{prob0}). As seen from
figure \ref{fig:4}, clear deviations of the extracted experimental
phase $\phi_e$ occur with regard to the reference phase $\phi$.

To understand how a smooth change in the atom-cavity detuning
$\Delta (t)$ affects the frequency and phase of the predicted
final-state probability, a number of steps have to be carried out
for both models of Sections 2.2 and 2.3: (i) to evaluate numerically
the final-state probability $P_{\tau}(T)$ or
$\widetilde{P}_{\tau}(T)$ for the time delay intervals $I_{\nu}$
from the experiment; (ii) to fit the numerical results from step (i)
to the analytical function (\ref{prob}) in order to extract the
cosine arguments for every time interval; and (iii) to repeat the
previous two steps for different values of the `switching' parameter
$\tau$. The (cosine) arguments obtained from this procedure can then
be compared to the experimental dots from figure \ref{fig:4} in
order to determine the parameter $\tau$ that agrees better with the
experiment.

\begin{figure}
\begin{center}
\includegraphics[width=0.7\textwidth]{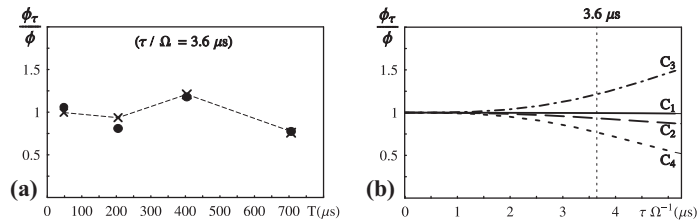}\\
\caption{Comparison of the relative phase $\frac{\phi_{\tau}}{\phi}$ for a
smooth change of the atom-cavity detuning (cf.\ Section 2.2) with experiment.
(a) Results are shown for the `switching' parameter
$\tau \Omega^{-1} = 3.6 \mics$, for which the best agreement with the
experimental data (dots) is found. The theoretical values (crosses) are
joined by dashed lines to guide the eyes of the reader. Figure (b) displays
the relative phase $\frac{\phi_{\tau}}{\phi}$ as a function of $\tau$
(curves $C_{\nu}$). Every curve corresponds to a phase $\phi_{\tau}$ as
obtained for the time intervals $I_{\nu}$. The vertical dashed line indicates
the $\tau$ value from plot (a).}
\label{fig:5}
\end{center}
\end{figure}

Figure \ref{fig:5} compares the relative phase $\phi_{\tau} / \phi$
for a smooth change of the atom-cavity detuning as modeled in
Section 2.2. In Figure \ref{fig:5}(a), results are shown for the
`switching' parameter $\tau / \Omega = 3.6 \mics$, for which the
best agreement with the experimental data (dots) is found. In
addition, figure \ref{fig:5}(b) displays the relative phase
$\frac{\phi_{\tau}}{\phi}$ as a function of $\tau$ for each time
intervals $I_{\nu}$. As seen from this Figure, our mixed
analytic-numerical treatment of the atom-cavity state gives (as
expected) the same final-state probability (\ref{prob0}) in the
limit $\tau \rightarrow 0$. Note that the differences between
$\omega_{\tau}$ and $\omega$ is about two orders of magnitude
smaller than the differences in the phases (and are therefore not
displayed here). This confirms that the frequency of the final-state
probability is less sensitive to `undesired' effects in the
atom-cavity interaction than the phase shift. The best agreement
between our predictions and experiment is found for a `switching
time' $\tau / \Omega = \overline{\tau} / \Omega = 3.6 \mics$ that is
far outside of what is expected for the switch of the atomic
transition frequency from one to the other cavity mode in the
experiment \cite{pra64}. Therefore, despite of its rather large
effect on the phase-shift of the final-state probability, the finite
switching time cannot explain the observations from the experiment
as taken alone.

\begin{figure}
\begin{center}
\includegraphics[width=0.7\textwidth]{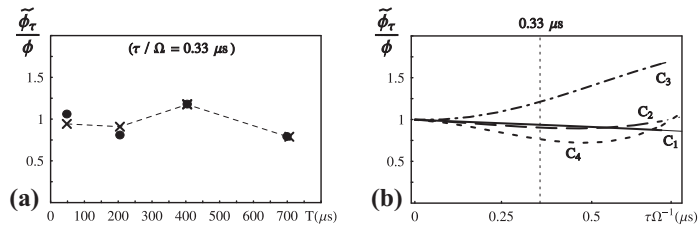}\\
\caption{The same as figure~\ref{fig:5} but for the
communication-channel model in Section 2.3,
$\frac{\widetilde{\phi}_{\tau}}{\phi}$. (a) Results are shown again
for the `switching' parameter $\tau \Omega^{-1} = 0.33 \mics$ with
the best agreement with the experimental data (dots). The
theoretical values (crosses) are joined by dashed lines to guide the
eyes of the reader. Figure (b) displays the relative phase
$\frac{\widetilde{\phi}_{\tau}}{\phi}$ as a function of $\tau$
(curves $C_{\nu}$). Every curve corresponds to a phase
$\widetilde{\phi}_{\tau}$ as obtained for the time intervals
$I_{\nu}$. The vertical dashed line indicates the $\tau$ value from
plot (a).}
\label{fig:6}
\end{center}
\end{figure}

In the communication channel model from Section 2.3, in contrast,
the atom interacts for a short period with both cavity modes
simultaneously, leading to an effective `communication' between the
cavity modes. Figure \ref{fig:6}(a) displays again the relative
phase $\widetilde{\phi}_{\tau} / \phi$ for that switching parameter,
$\tau / \Omega = 0.33 \mics$, for which the best agreement with the
experimental data (dots) is found. This value nicely fits to the
expected time that is needed to `switch' the atomic frequency
between the two cavity modes. Moreover, the relative phase
$\widetilde{\phi}_{\tau} / \phi$ as function of $\tau$ is shown in
figure \ref{fig:6}(b) for each time intervals $I_{\nu}$ separately.
We therefore conclude that a finite switch together with an
effective communication of the two cavity modes, i.e.\ the model of
the communication channel, is appropriate for describing the
atom-cavity interaction when the atom passes through the cavity.

Let us mention once more that numerical simulations were performed
only `while' the atomic frequency is switched between the cavity
modes ($\tau \Omega^{-1} \leq 1\:\mics$), both for the single-mode
(Section 2.2) as well as the communication-channel model (Section
2.3). For all other periods, the time evolution of the atom-cavity
state was followed analytically (\ref{sols0}) owing to a purely
resonant atom-cavity interaction. The slight deviation between the
predicted and observed phases [Figures \ref{fig:6}(a) and
\ref{fig:5}(a)] clearly indicate that there is no perfect agreement
with the experimental data. However, no error bars has been given on
the final-state probability in \cite{pra64}, and hence this slight
disagreement should be later on refined in accordance with error
bars which occur in cavity QED experiments.

\section{Summary and outlook}

The coherent evolution of the atom-cavity state in bimodal (cavity)
experiments has been analyzed for a realistic time-dependence of the
detuning of the atomic transition frequency. Apart from a `smooth
switch' of the atomic resonance from one to the other cavity mode
during a short but finite time interval (Section 2.2), we considered
also an additional (effective) interaction, a so-called
communication channel, between the two cavity modes (Section 2.3).
While, in the former case, the atom interacts at any time with just
one mode of the photon field, either resonantly or non-resonantly, a
simultaneous interaction with both field modes is modeled by means
of the communication channel. The outcome of the experiment is the
final-state probability for that the probe atom is observed in the
(excited) Rydberg state; this probability is analyzed by combining
the analytical solution of the Jaynes-Cummings Hamiltonian with
numerical simulations for all those time intervals during which the
atom-cavity interaction is tuned from one cavity mode to the other.

Comparison of our model computations have been made especially with
the recent measurements by Rauschenbeutel and coworkers
\cite{pra64}. To this end, a series of computations have been
carried out for the final-state probability as function of the delay
time $T$ and for different (realistic) choices of the `switching
time' $\tau$, the model parameter, during which the atom is
non-resonant to the cavity modes. Making use of typical cavity
parameters as reported in \cite{pra64}, it is demonstrated that the
agreement between the predicted and experimental phase can be
improved by allowing a `communication' between the two cavity modes,
cf.\ Figures~\ref{fig:5} and \ref{fig:6}. Together with the recent
study by de Magalhaes and Nemes \cite{pra70}, who investigated the
damping of the probability amplitude due to the coupling of the
cavity to the environment, the use of the communication channel from
above brings the theoretical predictions in good agreement with
experiment. For the interpretation of future bimodal experiments,
therefore, it seems appropriate to take into account both, the
decoherence as well as the details of the atom-cavity detuning
dynamics.

In fact, several proposals have been reported for performing bimodal
cavity experiments \cite{pra76, pla339, pra68, jmo51, jmo52, pra66,
pra67}. In the proposal by Zubairy \etal \cite{pra68}, for instance,
a bimodal cavity is utilized to realize a quantum phase gate in
which the two qubits are given by the cavity modes. Based on this
gate, the authors suggested a scheme that enables one to implement
Grover's search algorithm. Another fruitful branch of the bimodal
cavity applications represent proposals \cite{jmo51, jmo52, pra66,
pra67}, where the schemes for engineering of various entangled
states between the atomic and/or photonic qubits have been reported.
All these papers ignore, however, the effects of the
`non-instantaneous' detuning of the atom-cavity interaction from one
cavity mode to the other, as well as the effective interaction among
the cavity modes. Further investigation are therefore required to
better understand how well these schemes may work in practise.
Finally, we mention the papers \cite{pra76, pla339} where it has
been proven that the coupling of both cavity modes to a common
reservoir, as in line with $H_{coupled}$ (\ref{coupled}), induces
the tunneling of a field state from one cavity mode to another mode
of the same cavity, and thus, opens a way to implement the
environment assisted (short-distance) teleporting inside a bimodal
cavity. We note that in order to follow the time evolution of such
quantum systems embedded into a reservoir or under the external
noise and to analyze different (entanglement or separability)
measures, a `quantum simulator' has been developed recently in our
group \cite{Radtke:06} that can be utilized for such studies in the
future.

\ack We acknowledge the fruitful discussion with Thomas Radtke. This
work was supported by the Deutsche Forschungsgemeinschaft (DFG).

\end{document}